\documentclass{epl}

\title{Transition to a Superconductor with Insulating Cavities}
\author{Mauro M. Doria \and Antonio R. de C. Romaguera}
\institute{Instituto de F\'{\i}sica, Universidade Federal do Rio de Janeiro,
C.P. 68528, 21941-972, Rio de Janeiro RJ, Brasil}

\begin{document}
\shorttitle{\small Transition to a Superconductor with Insulating Cavities}
\shortauthor{\small Doria and Romaguera}
\maketitle

\begin{abstract}

An extreme type II superconductor with internal insulating regions, namely cavities, is studied here.
We find that the cavity-bearing superconductor has lower energy than the defect-free superconductor above a critical magnetic induction $B^*$ for insulating cavities but not for metallic ones.
Using a numerical approach for the Ginzburg-Landau theory we compute and compare free energy densities
for several cavity radii and at least for two cavity densities, assuming a cubic lattice of spherical cavities.

\end{abstract}

The interface between the superconducting state and an exterior medium has a delicate energetic balance whose importance was  appreciated by Abrikosov\cite{Ab57} in his seminal work of 1957 that predicted vortices in superconductivity.
The superconducting density decays near this interface with an energy cost per area of $\xi H_c^2/8\pi$, $\xi$ is the coherence length, $H_c^2/8\pi$ the condensate energy, and  $H_c$ the superconductor's critical field.
The external applied field penetrates inside the superconducting state with an energetic cost opposite to the previous one, $-\lambda H_c^2/8\pi$, $\lambda$ being the London penetration length.
The addition of these two energies gives the energetic cost of the compound's external physical surface and also determines the nucleation of domain walls inside this compound, which becomes a sum of superconducting and non superconducting regions.
According to the above qualitative argument this nucleation is possible for a compound with a Ginzburg-Landau parameter $\kappa = \lambda/\xi$ larger than one, which is a type II superconductors.
As pointed out by Yu. N. Ovchinnikov\cite{O80} long ago, the investigation of various types of inclusions in superconducting materials is of particular interest.
The question here is which inclusions can be considered as domain walls, spontaneously nucleated inside the superconductor by energetic reasons.

Let us consider here an extreme type II superconductor ($\kappa >> 1$) with non-superconducting regions in its interior, that we call cavities, with typical size of $\xi$, and separated by a dozen of $\xi$.
The fact that cavities introduce novel properties to the superconductor has been previously shown by Doria and Zebende\cite{DG02}.
However cavities can be insulating or metallic and this has important consequences for the properties of  the cavity-bearing superconductor.
Here we show the remarkable property that only insulating cavities can turn the superconductor into some kind of bubble system, that is, turn its coexistence  with non-superconducting regions into a stable phase.
For both cases the cavity-bearing superconductor share similar properties, such as multiple trapping of vortices by a single defect\cite{B92}, meta-stability near transition fields\cite{YP00}, and giant vortex states.
Internal surface superconductivity above the upper critical field is only possible for insulating cavities\cite{DS99,DG02}.
Here we show that only for insulating cavities, but not for metallic ones, the cavity-bearing superconductor presents  a subtle change of energetic balance, described by the difference in free energy density,
\begin{eqnarray}
\delta F = F_{defect-free} - F_{cavity}
\end{eqnarray}
between the  cavity-bearing and the defect-free superconductors.
This difference $\delta F$ changes sign above a critical magnetic induction $B^{*}$, located  near and below the upper critical field.
The metallic cavity-bearing superconductor always has larger energy than the defect-free one, $\delta F < 0$, though it has  more superconducting volume than  the insulating one because of the proximity effect inside the cavity.
Somehow the property that defines the insulating cavity, namely that the supercurrent component normal to the cavity surface vanishes\cite{SJDG63}, plays an important role in this energetic balance.
In conclusion $\delta F > 0$ is only possible for insulating cavities above some critical density $B^{*}$, whose properties, as a function of cavity size, are determined here in generalization of the work of Doria and Zebende \cite{DG02}.
The present study can be of relevance for fabricated hollow three-dimensional superconductors.
Artificially made superconducting films\cite{MO96} with a regular array of open holes\cite{BBP94}, and of blind holes\cite{BP96}, have been constructed in the past.

In this letter we study the critical magnetic induction $B^{*}$ on a cavity-bearing superconductor modeled by spheres of radius $R$ equal or larger than $\xi$ that form a periodic cubic lattice, and obtain its properties, such as its dependence on the cavity radius $R$, and on the cavity density $1/L^3$ for insulating cavities, $L$ being the distance between two consecutive cavities on the cubic lattice.
We find that the difference in free energy density can reach $10^{-3}H_{c}^2/4\pi$ for the typical lattice  $L = 12.0\xi$ of insulating cavities treated here.
The present results support the view of insulating cavities as domain walls spontaneously nucleated inside the superconductor by energetic reasons.

A cavity can trap many vortices simultaneously in its interior.
The number of trapped vortices increase with the magnetic induction and eventually reaches a saturation limit.
Near the upper critical field the vortex repulsion weakens and the pressure exerted by the external vortices into the trapped ones is strong to produce a giant vortex inside the cavity.
The process of trapping vortices by a cavity occurs similarly to the Mkrtchyan and Shmidt \cite{MS72,O80} capture process, who found long ago the saturation limit of $R/2\xi$, for a columnar defect of radius $R$.
To understand it, consider an empty cavity which sets  an attractive potential to an external vortex.
Once captured the newly formed vortex-cavity system sets a new potential barrier for another external vortex, which is repulsive, far from the cavity and attractive, close to it.
The capture process can go on until a maximum number of vortices inside the cavity is reached, which is the saturation limit of the cavity.
When the external vortices meet the energy requirement for the entrance of a new vortex into the cavity the system becomes meta-stable.
Cavities, like misoriented columnar defects, can cause the vortex lines to intersect inside but not outside them, where vortex-vortex repulsion becomes dominant.

The numerical search for the free energy minima is done for a fixed vortex density, namely for $N$ fluxons
piercing two parallel faces of the cubic unit cell that corresponds to a magnetic induction $B = 2 \pi \kappa N (\xi/L)^2$.
Because  screening currents are not being included here, the local magnetic field is constant and equal to the magnetic induction everywhere.
The unit cell is the simplest possible one, that is, with a single cavity in its center, in order that the numerical calculation time becomes feasible.
To describe it, a mesh of $P^3$ points is used, and for the present simulations, the distance between two consecutive mesh points is $2\xi/3$, such that holds that $L=2\xi(P-1)/3$.
A gauge invariant\cite{DGR90} modified version of the Ginzburg-Landau theory is used here.
It treats superconducting and non-superconducting regions on equal footing \cite{DS99,DG02} because it effectively incorporates the appropriate boundary conditions  into the free energy functional.
The theory is numerically solved in this mesh through the Simulated Annealing method\cite{DGR89}, a Monte Carlo thermal procedure.
\begin{figure}
\centerline{\includegraphics[scale=0.79]{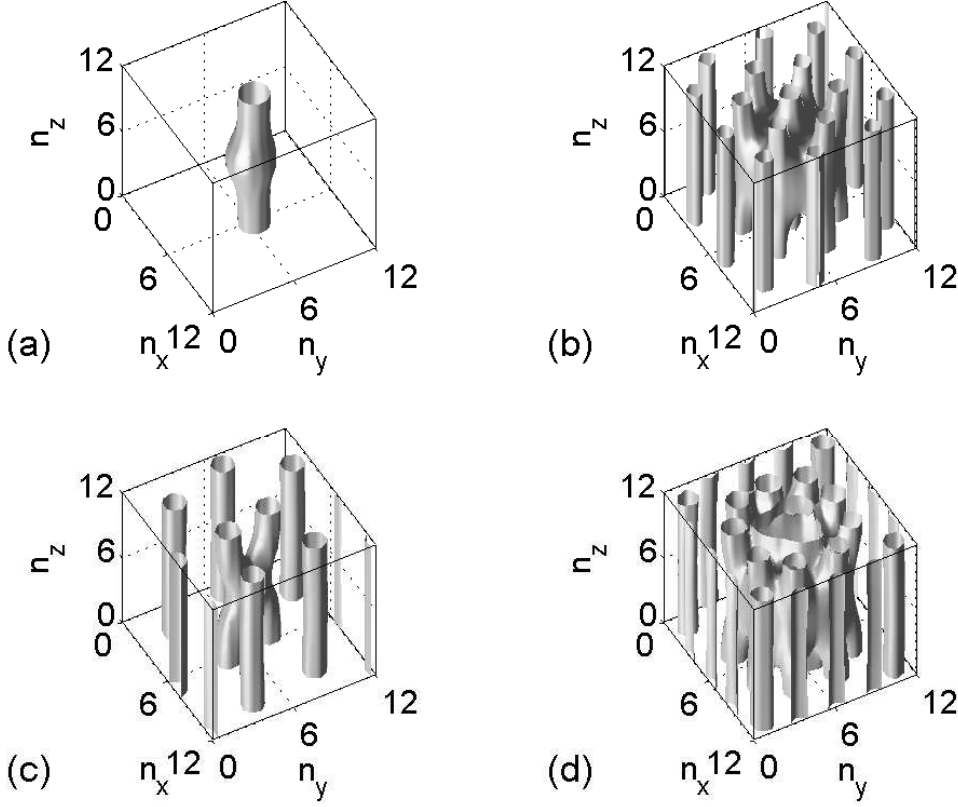}}
\caption{Iso-surface plots  are shown here for the following $L=12.0\xi$ cavity lattices:
(a) $|\Delta|_{iso}^2=0.3365$, $|\Delta|_{max}^2=0.9991$,  metallic cavity radius $R=2.0\xi$, and $N=1$ vortices in the unit cell ($\delta F < 0$);
(b) $|\Delta|_{iso}^2=0.1418$, $|\Delta|_{max}^2=0.4253$, metallic cavity radius $R=3.0\xi$, and $N=18$ vortices in the unit cell ($\delta F < 0$);
(c) $|\Delta|_{iso}^2=0.2823$, $|\Delta|_{max}^2=0.8353$, insulating cavity radius $R=2.0\xi$, and $N=9$ vortices in the unit cell ($\delta F < 0$);
(d) $|\Delta|_{iso}^2=0.1079$, $|\Delta|_{max}^2=0.3235$, insulating cavity radius $R=4.2\xi$,  and $N=22$ vortices in the unit cell ($\delta F > 0$)}
\label{iso4}
\end{figure}
Fig.~\ref{iso4} shows three-dimensional plots of the superconducting density in the unit cell for some selected configurations, all of them obtained using $19^3$ mesh points to describe the cubic unit cell with side equal to $12.0\xi$.
Fig.~\ref{iso4}(a) shows a $R=2.0\xi$ metallic cavity with $N=1$.
Fig.~\ref{iso4}(b) shows a $R=3.0\xi$ metallic cavity with $N=18$.
It contains 6 vortices forming a central ring surrounding the cavity with 1 trapped vortex.
Fig.~\ref{iso4}(c) shows a $R=2.0\xi$ insulating cavity with $N=9$.
It contains two trapped vortices inside the cavity.
Fig.~\ref{iso4}(d) shows a $R=4.2\xi$ insulating cavity with $N=22$.
It contains 9 vortices forming a central ring surrounding the cavity with 3 trapped vortices about to collapse into a single giant vortex.
Figs.~\ref{iso4}(a),(b), and (c) are $\delta F < 0$ configurations whereas Fig.\ref{iso4}(d) is the only $\delta F > 0$ configuration.
For high vortex density the repulsion among vortices is weak enough that clustering around the defect becomes possible.
This can be seen in both figs.~\ref{iso4}(b) and (d). However only for fig.~\ref{iso4}(d) the energy is smaller than the corresponding defect-free superconductor.

The cavity is described by a step-like function, zero inside the cavity and one in the superconducting region,
made smooth for numerical reasons, $\tau({\vec x}) = 1 - 2 / \lbrace \; 1 + \exp{{\lbrack(|{ \vec x} |/R)}^K \rbrack}\;\rbrace $, with $K=8$.
The free energy density is,
\begin{eqnarray}
F = \int {{dv}\over{V}} \; \tau' \xi^2  |({\vec \nabla} - {{2\pi i}\over{\Phi_0}} {\vec A})\Delta|^2
-\tau |\Delta|^2 + {1 \over 2} |\Delta|^4, \label{eq:glth}
\end{eqnarray}
expressed in units of the critical field energy density, $ H_c^2/4\pi $ and  the superconducting density $|\Delta|^2$ normalized between zero and one.
The metallic cavity is defined by $\tau'=1$ since this condition allows the condensate to exist as a fluctuation outside the superconductor, whereas for the insulating cavity it cannot since $\tau'=\tau$\cite{DS99,DG02}.
The defect-free superconductor corresponds to $\tau = 1$ everywhere.
\begin{figure}
\centerline{\includegraphics[scale=0.85]{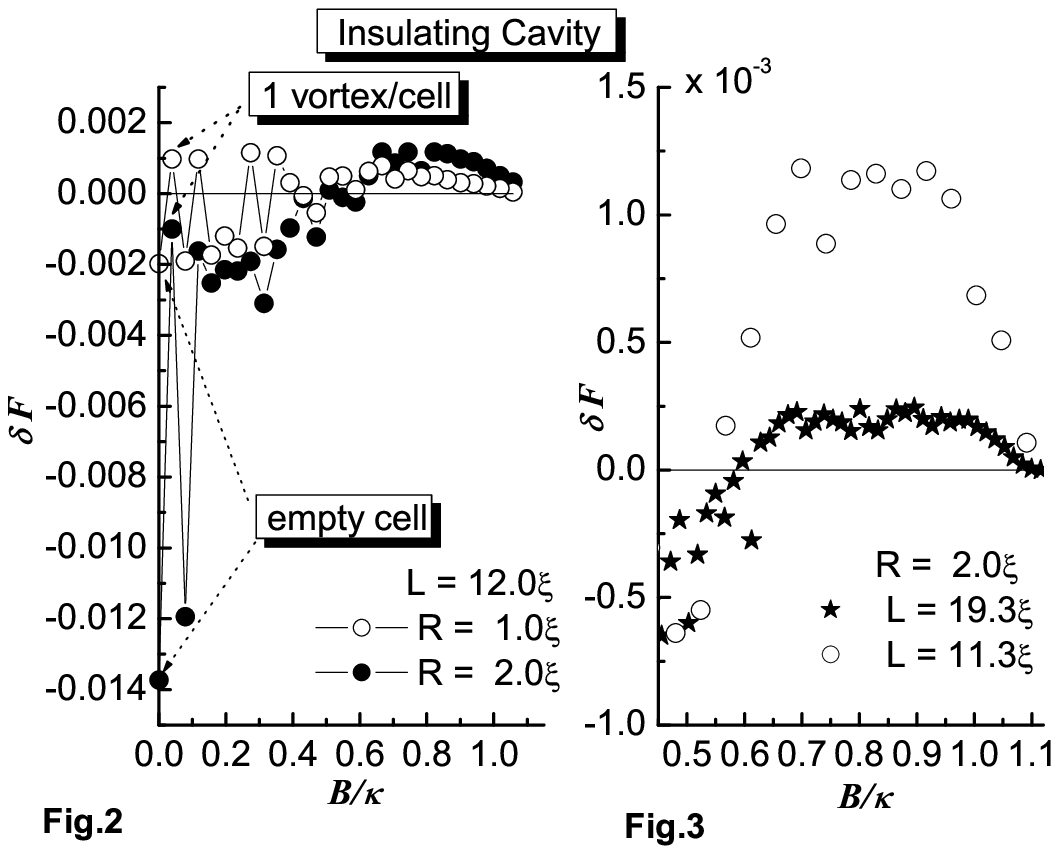}}
\caption{The free energy density difference between a superconductor with a cubic lattice ($L=12.0\xi$) of insulating cavities and the defect-free superconductor is shown here versus the magnetic induction $B$ for two cases of cavity radii, $R=1.0\xi$ and $2.0\xi$.}
\label{insr1e2}
\caption{The free energy density difference between a superconductor with a cubic lattice of insulating cavities and the defect-free superconductor is shown here versus the magnetic induction $B$ for a single kind of cavity ($R=2.0\xi$), and two cases of cavity lattices, $L=11.3\xi$ and $19.3\xi$.}
\label{insn18e30}
\end{figure}
To understand the change of sign in $\delta F$ let us start with the simplest possible case of no vortices in the unit cell ($B=0$).
In this case the defect-free superconductor has maximum density everywhere, $|\Delta|^2=1$,  and its free energy density is  simply $F_{defect-free}= -0.5$.
The cavity-bearing superconductor has higher energy than the defect-free one, as numerically determined
for insulating cavities ($L = 12.0\xi$): $F_{cavity} =$ $-0.498$, $-0.486$,$-0.456$,$-0.319$,$-0.306$, and $-0.1918$, for $R=1.0\xi$ to $6.0\xi$, respectively, and this is fairly well described by $F_{cavity}\approx (1-4\pi R^3/3L^3)F_{defect-free}$.
Inside the cavity the order parameter vanishes, $|\Delta|^2=0$, rendering the free energy of the cavity-bearing superconductor approximately  equal to the defect-free case removed of the cavity volume.
However there is also the kinetic energy contribution to $F_{cavity}$, caused by the curvature of the order parameter near the cavity surface, an effect that becomes more pronounced for large cavities.
The case of one vortex in the unit cell is also worth of discussion.
The vortex nucleates over the cavity to take advantage of the existing non-superconducting region to minimize its total core energy cost.
For the sake of the argument let us assume that the vortex core is a non-superconducting cylinder of cross section radius equal to $\xi$, which implies that for $R \le \xi$ the cavity is fully inside the vortex core, but not for $R > \xi$.
In this respect fig.~\ref{iso4}(a) could be regarded as pictorial view  for $R > \xi$ showing that there is an extra cavity volume outside the vortex core which has some energy cost.
In conclusion the cavity-bearing superconductor should have higher energy than the defect-free one only if $R > \xi$ and
indeed, as shown in fig.~\ref{insr1e2}, $\delta F$ is positive for $R =1.0\xi$, and negative for $R = 2.0\xi$.
For more than two vortices in the unit cell, several competing effects contribute to determine $\delta F$.
Multiple trapping by the cavity  decreases the total energy nucleation of the vortices since the cavity volume is simultaneously occupied by more than one vortex.
The cavity works as an efficient vortex ``crossroad" and lowers the total sum of vortex self-energies.
From the other side the  mutual vortex repulsion away from the cavity introduces a curvature to the vortices of opposite energetic cost, since vortices become longer as compared to the straight vortex lines found in the  defect-free superconductor.
Thus above the magnetic induction  $B^{*}$  the energetic balance is in favor of the cavity-bearing superconductor in case of insulating cavities.

The difference  $\delta F$ versus $B$ is plotted in fig.~\ref{insr1e2} for two kinds of distinct radii, namely $R=1.0\xi$ and $2.0\xi$, both forming a $L=12.0\xi$ insulating cubic lattice.
For the $R=1.0\xi$ cavity one expects $\delta F > 0$ for any $B$ because the cavity fully fits inside the vortex core and lowers the vortex energy nucleation with no extra cost to the superconducting state.
However an oscillatory behavior is observed for low density that disappears in the high limit, above $B = 0.5\kappa$ where $\delta F >0$.
All the $\delta F < 0$ $R=1.0\xi$ points of fig.~\ref{insr1e2} are empty cavity configurations.
This is a consequence of weak pinning by the cavity and strong vortex-vortex repulsion, such that for certain densities, vortex trapping by the cavity is less important than an efficient vortex arrangement inside the unit cell to minimize the overall repulsive interaction.
Thus the $\delta F < 0$ $R=1.0\xi$ points are not truly $\delta F < 0$ states, just a consequence of our choice of unit cell. These states flip to $\delta F > 0$, provided that a larger unit cell is able to accommodate the repulsion among vortices and have all the cavities filled with trapped vortices.
For the $R=2.0\xi$ cavity the overlap is such that part of the cavity remains external to the vortex core and this adds energy to $F_{cavity}$, resulting in a $\delta F < 0$ state.
Two vortices inside unit cell have commensurability problems since the vortex-vortex repulsion is strong enough to overcome the attraction to the cavity and this results in both vortices out of the cavity not only for $R=1.0\xi$ but also for $R=2.0\xi$.
The energetic balance is subtle for more than two vortices in the unit cell, as shown in fig.~\ref{iso4}.
The outcome of fig.~\ref{insr1e2} is that for both raddi the phase $\delta F > 0$ only becomes stable for an induction larger than $B^{*} \approx 0.6\kappa$.

Let us consider the problem of a superconductor with just one insulating cavity, which is reached in the limit $L \rightarrow \infty$ of a lattice system.
From the present lattice results we find evidence that the one cavity superconductor has lower energy than defect-free one above some critical magnetic induction.
To reach this conclusion we consider different cavity densities, namely, $L=11.3\xi$ and $L'=19.3\xi$, for a single kind of cavity, $R=2.0\xi$.
Fig.~\ref{insn18e30} shows that the $\delta F>0$ plateaus scale inversely proportional to the unit cell volume, $L'^3 \delta F' \approx L^3 \delta F$ since ($\delta F'\approx 0.24 \times 10^{-3}$, $\delta F \approx 1.2 \times 10^{-3}$), and the volume ratio is $(L/L')^3\approx 0.2$. Also notice that  $B^{*} \approx B'^{*}\approx 0.55 \kappa $.

\begin{figure}
\centerline{\includegraphics[scale=0.80]{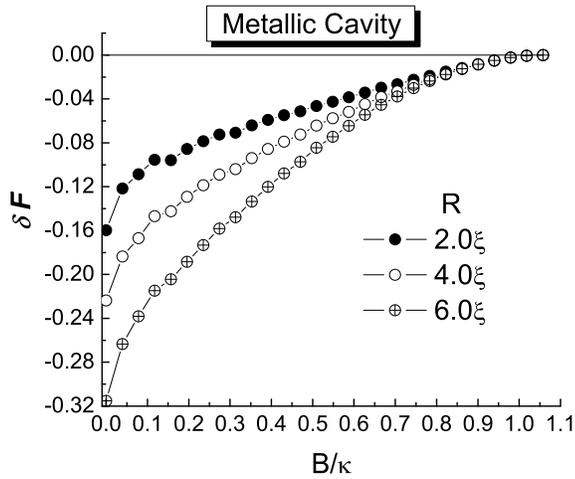}}
\caption{The free energy density difference between a superconductor with a cubic lattice ($L=12.0\xi$) of metallic cavities and the defect-free superconductor is shown here versus the magnetic induction $B$ for cavity radii $R=2.0\xi$, $4.0\xi$, and $6.0\xi$. The inequality $\delta F < 0$ is always satisfied for all radii.}
\label{metradii}
\end{figure}

\begin{figure}
\centerline{\includegraphics[scale=0.85]{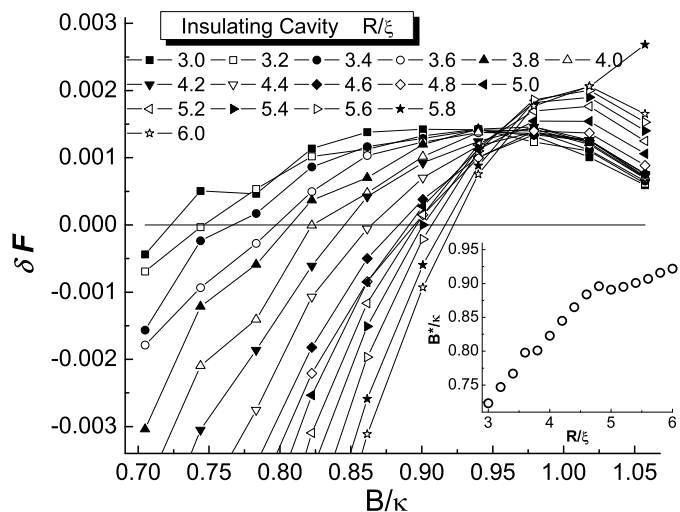}}
\caption{The free energy density difference between a superconductor with a cubic lattice ($L=12.0\xi$) of insulating cavities and the defect-free superconductor is shown here versus the magnetic induction $B$ for several cavity radii, ranging from $R=3.0\xi$ to $6.0\xi$, every $0.2\xi$.
The plot is focused in the region where this energy difference flips sign and the corresponding induction where this occurs is plotted versus R in the inset. }
\label{insradii}
\end{figure}

The metallic cavity-bearing superconductor is studied in fig.~\ref{metradii}, which shows  $\delta F$ versus $B$ for three cavity radii, $R=2.0\xi$, $4.0\xi$, and $6.0\xi$ and all the data points satisfy $\delta F < 0$.
In case of insulating cavities for a certain magnetic induction the largest possible cavity that renders  $\delta F > 0$ can be read off from fig.~\ref{insradii}, which shows several sets of $(\delta F,B)$ data points, each set having its points connected by straight lines. The sets are associated to distinct cavity radii, ranging from $R=3.0\xi$ to $6.0\xi$.
These data points do not form smooth curves because of the many metastable configurations whose energies are very near in value.
A large cavity takes away a large volume that otherwise would be in the superconducting state.
It also demands a great amount of surface energy to nucleate, because of the curvature of the order parameter near the interface.
Consequently all the $\delta F$ curves steeply dip into the negative region for low $B$, as indicated
by fig.~\ref{insradii}.
But for $B > 0.7\kappa$ the situation changes and a critical induction $B^{*}$ exists for each cavity radius.
$B^{*}$ is approximately obtained as the intersection of the zero axis with the straight line connecting the two points immediately above and below the zero axis.
$B^{*}$ approaches  the upper critical induction $B_{c2}=\kappa$ for very large cavities because the superconductor is already near the normal state and the space left for the superconducting state outside the cavities shrinks to zero.
The vortex states above $B_{c2}$ are surface superconductivity states\cite{DG02}.
The inset of fig.~\ref{insradii} shows $B^{*}$ obtained by this approximate method versus $R$.
From this figure we conclude that for a given induction there are many cavity-bearing superconductors with less energy than the defect-free one up to a maximum possible cavity radius. Obviously this maximum radius is also cavity density dependent since above a certain radius $R \ge L/2$ the cavities touch each other.

In summary we have studied a superconductor with a cubic lattice of spherical cavities and compared its free energy with a defect-free superconductor.
For high vortex density, as compared to the cavity density, properties such as multiple trapping of vortices by a single cavity and vortex curvature play an important role, providing a substantially different view of the superconductor from the standard picture.
We found here that the spontaneous nucleation of insulating cavities as domain walls is a possible process to occur inside a superconducting compound above some critical field $B^{*}$.

\acknowledgments
Research supported in part by Instituto do Mil\^enio de Nano-Ci\^encias, CNPq, and FAPERJ (Brazil).

\end{document}